\documentclass[fleqn,10pt]{wlscirep} 
\title{Trajectory stability in the traveling salesman problem
}
\usepackage[draft,inline,nomargin]{fixme}
\fxsetup{theme=color}
\usepackage[normalem]{ulem} 
\newcommand{\unam}{Universidad Nacional Aut\'onoma de M\'exico, 01000 M\'exico D.F., Mexico}
\renewcommand{\paragraph}[1]{{\it #1.--}}

\newcommand{\ifunam}{Instituto de F\'{\i}sica, \unam}

\newcommand{\fref}[1]{Figure \ref{#1}}
\newcommand{\eref}[1]{Eq.~(\ref{#1})}

\newcommand{\iimas}{Instituto de Investigaciones en Matem\'aticas Aplicadas y en Sistemas, \unam} 
\newcommand{\ccc}{Centro de Ciencias de la Complejidad, \unam}
\newcommand{\aalto}{Department of Computer Science, Aalto University School of Science, 00076 Aalto, Finland}
\author[1]{Sergio S\'anchez}
\author[1]{Germinal Cocho}
\author[1]{Jorge Flores}
\author[2,3,4,5]{Carlos Gershenson}
\author[2,6]{Gerardo I\~niguez}
\author[1,7,*]{Carlos Pineda}
\affil[1]{\ifunam}
\affil[2]{\iimas}
\affil[3]{\ccc}
\affil[4]{SENSEable City Lab, Massachusetts Institute of Technology, 02139 Cambridge, MA, USA}
\affil[5]{ITMO University, 199034 St. Petersburg, Russian Federation}
\affil[6]{\aalto}
\affil[7]{Faculty of Physics, University of Vienna, 1090 Wien, Austria}
\affil[*]{Corresponding author email: carlospgmat03@gmail.com}
\graphicspath{{images/},{figures/},{eps/}}

\newcommand{\sTSP}{bTSP}
\newcommand{\rTSP}{rTSP}

\begin{abstract} 
Two generalizations of the traveling salesman problem in which sites change their position in time are presented. The way the rank of different trajectory lengths changes in time is studied using the rank diversity. We analyze the statistical properties of rank distributions and rank dynamics and give evidence that the shortest and longest trajectories are more predictable and robust to change, that is, more stable.
\end{abstract} 
\begin{document}
\flushbottom
\maketitle

\thispagestyle{empty}

\section*{Introduction} 

Imagine that a certain product must be delivered, in the most efficient way, to
$N$ stores in a region. When stores are fixed at given sites in space,
finding the shortest path that links them all is the target of the {\it traveling salesman problem}
(TSP)~\cite{lawler1985traveling,gutin2007traveling}, which here we refer to as static TSP. But what happens if the product must be delivered to moving targets such as peddlers? The TSP becomes time dependent. Another variation of the TSP, related to
the peddlers' scenario, occurs when one or more sites disappear and then
reappear at other locations. Imagine, for example, a monkey looking for
fruits that grow in a set of trees. 
What happens if one of the trees ceases producing fruits? The monkey,
then, has to look for another tree. Time-dependent TSPs of this sort have several real-world applications,
such as vehicle routing~\cite{malandraki1992time} and
machine scheduling problems~\cite{Bigras2008685}.

\begin{figure} 
  \begin{center}
    \includegraphics{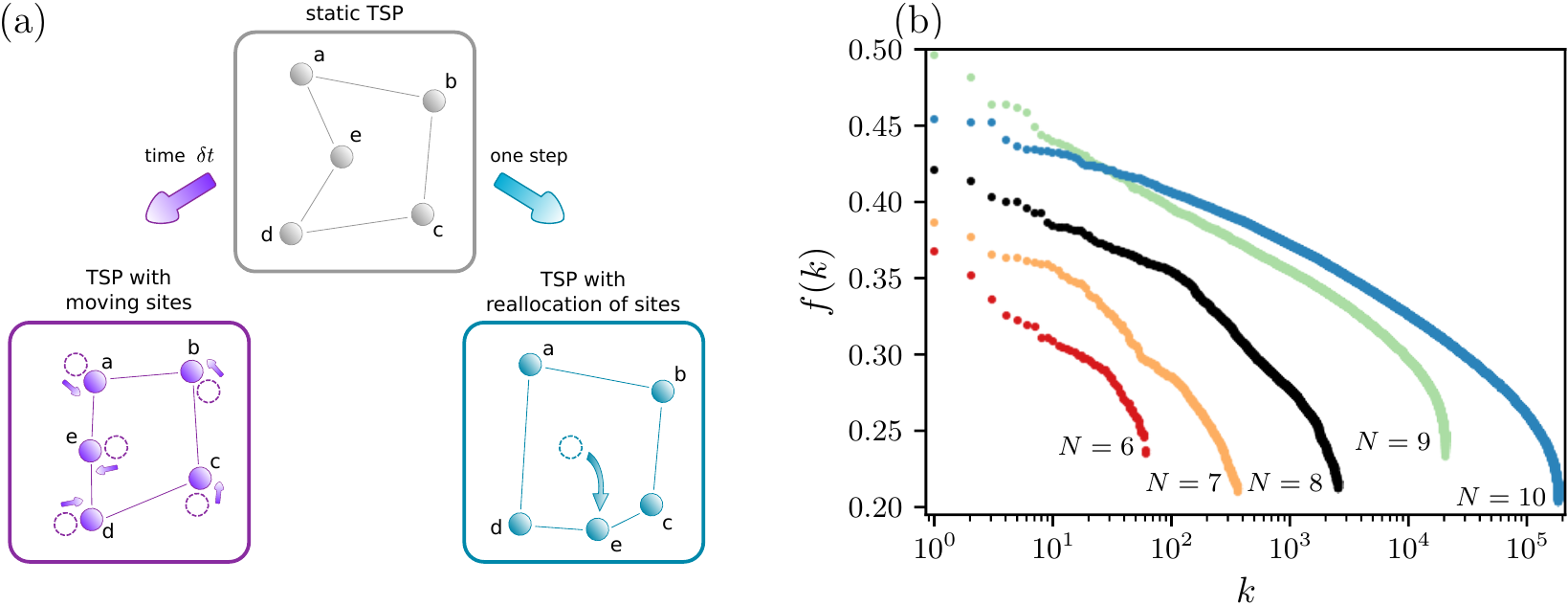}
   \end{center}   
   \caption{ {\bf Time-dependent traveling salesman problem with moving sites and  with reallocation
       of sites. Rank distribution for the TSP.} (a) Diagram showing the two
generalizations of the TSP considered here.  By allowing nodes to shift in
time, the static TSP (center) becomes time dependent.  In the TSP with
  moving sites (left), these are displaced in a small interval $\delta t$. In the
TSP with reallocation of sites (right), they are displaced to random positions
after one time step. In these TSP variations, nodes do not move during path traversals, that is,
the optimization problem of finding paths is solved for a static situation. Shortest paths are shown with
lines between sites.
(b) Rank distribution according to the inverse path length $f(k)$, for $N=6$ to
$N=10$ sites in particular but representative instances of the TSP.}
  \label{fig:gerardo}
\end{figure} 

Many variations of the TSP have been analyzed in recent
decades~\cite{lawler1985traveling,gutin2007traveling}. 
Previous research related to the TSP has focused mainly on producing algorithms to find shortest paths but, to our knowledge, the properties of longer
trajectories have not been discussed. In the present work we study the statistical properties  of all trajectories in two
generalizations of the TSP with time-dependent sites: the TSP {\it with moving sites} (bTSP), where sites can be interpreted as `boats' that move
gradually in a region, and the TSP {\it with reallocation of sites} (rTSP), where sites
move discontinuously across space [\fref{fig:gerardo}(a)].
In the peddlers' example above, peddlers might not move during one day (so
  trajectories do not change) but on the next they may have a different position (possibly
  modifying the shortest path, such that a new optimization process is needed to find
  it). That would be equivalent to assume that trajectory changes in time-dependent TSPs
  occur at a much slower time scale than the traversal of the paths by salesmen. If we
  rank trajectories by their length, we can analyze how the properties of this ranking
  change in time with measures commonly used in the study of hierarchy dynamics in complex
  systems, such as the rank distribution $f(k)$ and the rank diversity $d(k)$~\cite{cocho:idiomas,cochodos}.

Since the rank distribution was popularized in the 1930s by
Zipf~\cite{zipf}, it has been used to characterize complex
systems of different nature~\cite{clauset2009power}.  
We have recently discussed the cases of six
indo-european languages~\cite{cocho:idiomas} and of six sports and
games~\cite{cochodos}, where we found that $f(k)$ does not follow  Zipf's law and
varies slightly across cases. For all of these complex systems, we also studied how ranks
change in time by means of the rank diversity $d(k)$. Explicitly,
$d(k)$ is the number of different elements that have rank $k$ within a given
period of time $T$. In the complex systems we have studied up until now, the rank
diversity can be well approximated by a sigmoid curve.

\section*{Results} 
\subsection*{Static and time-dependent TSPs} 
Consider the static TSP with $N$ sites. 
We shall label each site by a, b, c, and so on. 
A {\it trajectory} is a closed path on these sites, so each trajectory can
characterized by a non-unique string of site labels. For example, the shortest
trajectory in \fref{fig:gerardo}(a) for the static TSP (top) is
``aedcb''. Note that one could have started the string at any site, and
even
changed the direction, so the same trajectory could have been labeled ``cdeab''.
Starting from his home city the salesman visits each site only once
and returns home, that is, he follows a trajectory. There are $(N-1)!/2$ different
trajectories, and the usual problem is to find the shortest one within this
 set. In this work we go further and rank each trajectory according to its length, so the shortest one
has the highest rank ($k=1$), and the longest has the lowest
rank ($k=k_{\max} = (N-1)!/2$). To study the rank distribution of the system,
$f(k)$ is taken as the inverse of trajectory 
length.  The rank distribution  of TSP
trajectories is presented in \fref{fig:gerardo}(b) for several random configurations
corresponding to different values of $N$. The results differ from Zipf's law,
but for all cases they have a similar shape, which resembles a beta distribution.

We now study the problem of stability of trajectories in the TSP.
Suppose the location of sites varies slowly over time, so that
the salesman can assume a static scenario for each travel,
but the shortest trajectory, and in fact the rank of all trajectories,
might change if the configuration of sites is modified enough. A toy model that captures this situation is the bTSP. 
Assume all $N$ sites are allowed to move within a $1\times1$ square as 
if they were boats. In the TSP with
moving sites, the $X$ and $Y$ components of the velocity of each site are random with a uniform
distribution between $\pm 1$.  The boats move with a constant velocity until they
reach a confining wall, where they bounce elastically [see \fref{fig:gerardo}(a)
and accompanying video in Supplementary Information].


\begin{figure} 
  \begin{center}
    \includegraphics[scale=0.9]{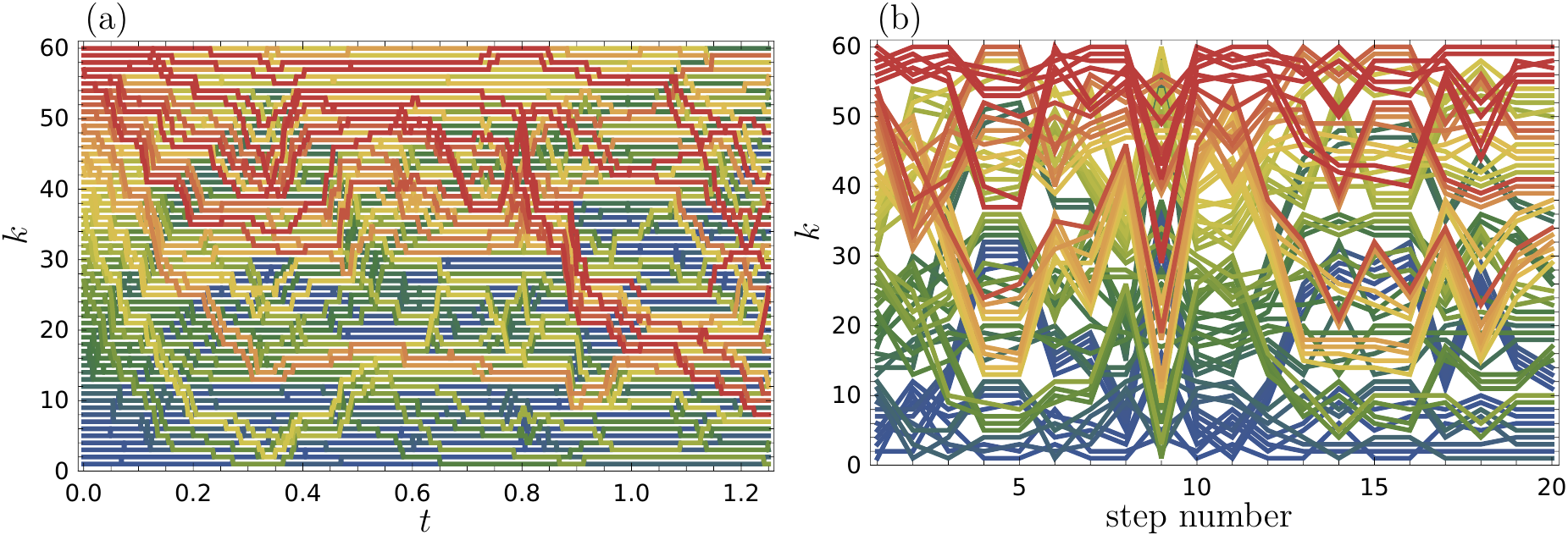}
  \end{center}
   \caption{
{\bf Time dependence of trajectory ranks.} 
Time dependence of trajectory ranks for the models depicted in \fref{fig:gerardo}(a), where particular
instances of $N=6$ are considered. Colors are an aid to see how
initial trajectory ranks diffuse over the $k$ axis as time goes by.
We show results for the (a) \sTSP{}, where trajectories spread slowly in 
time, and for the (b) TSP with reallocation of sites [\rTSP{}], where there are drastic changes from one time step to another, as well as a tendency for keeping initially shorter/longer trajectories short/long.
}
\label{fig:spaghetti}
\end{figure} 

Let us consider a different time-dependent perturbation of the TSP, the rTSP. Here $N$ sites are located in the unit
square with a uniform random probability, and all trajectories are ranked as explained above. 
This is the ``base'' configuration. Then one site is chosen at random and
reallocated to a random position in the unit square, after which trajectory ranks are calculated again [\fref{fig:gerardo}(a)]. 
After several iterations, we can explore this time-dependent process with the rank diversity. Even though a continuous time
dynamics does not exist as in the bTSP, we can still ask how many different trajectories occupy a given rank $k$, so the rank diversity is well defined.

\subsection*{Temporal evolution of trajectory ranks} 

We first explore how trajectory ranks change with time in both
the bTSP and rTSP (\fref{fig:spaghetti}). 
For the bTSP, trajectories initially at the extremes (i.e. high or low
$k$) tend to remain in place, until at some point they detach and explore
a wider range of ranks [\fref{fig:spaghetti}(a)]. 
That is, in the edges of the plot (high and low $k$), rank as a function of $k$ tends to
be a horizontal straight
line, whereas for middle $k$ the corresponding lines vary more. 
For the rTSP a slightly different behavior is observed [\fref{fig:spaghetti}(b)]. Here we note that time is an auxiliary variable with no relevant meaning, since we are
simply selecting random positions to relocate randomly chosen sites. As these positions are taken as a series of random and uncorrelated values, 
reordering such a sequence in time makes no statistical difference. Thus, it is reasonable to expect that trajectory ranks for the
rTSP vary more than for the bTSP. Despite this fact, we also observe that the
trajectories near the extremes tend to keep the same values of $k$, while intermediate trajectories do not exhibit such regularity, just like in the bTSP.

\begin{figure} 
  \begin{center}
    \includegraphics{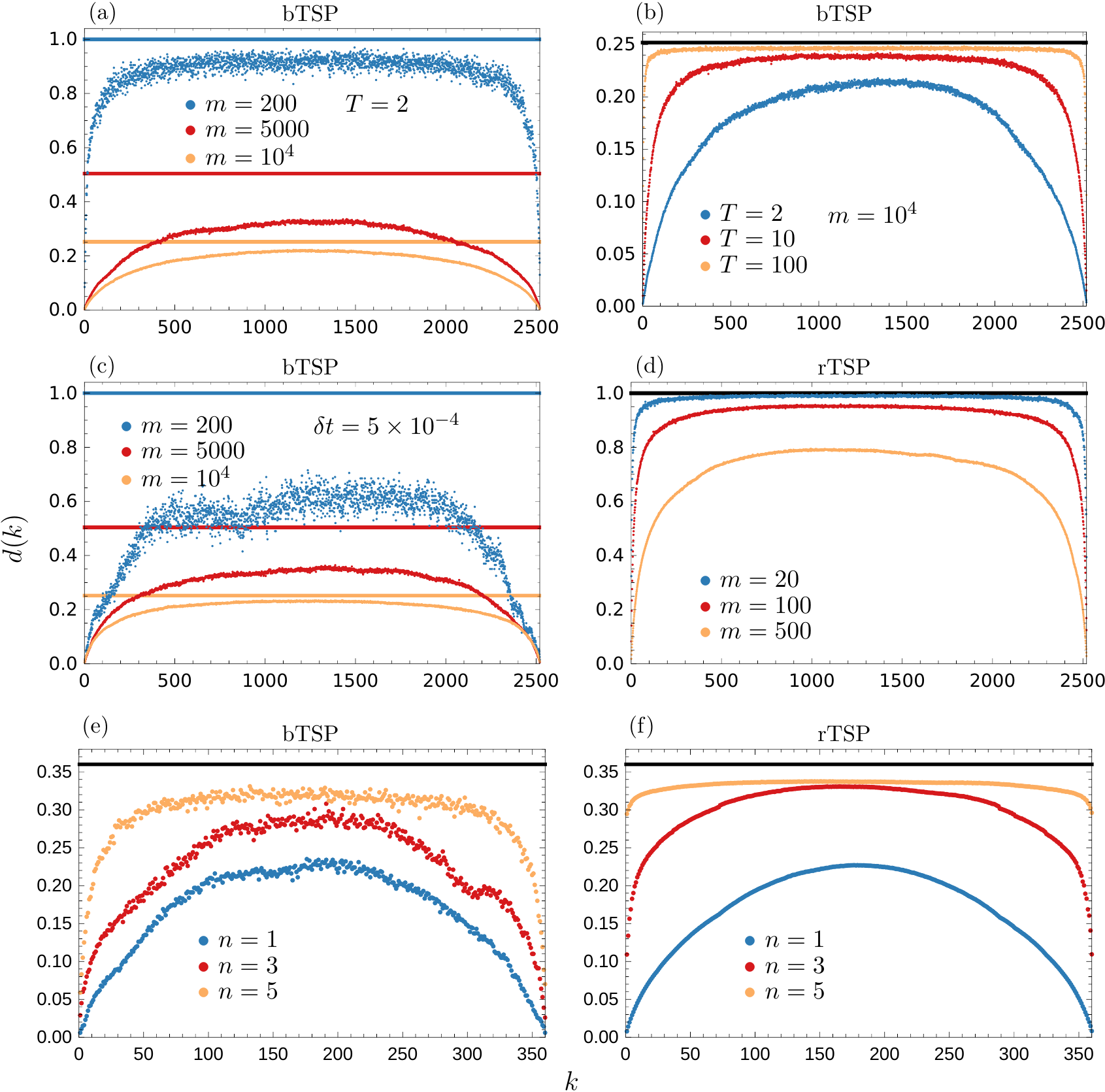}
  \end{center}
  \caption{ 
{\bf Parameter dependence of rank diversity.} We analyze how
changing parameter values affects the rank diversity for two time-dependent versions of the TSP. 
For panels (a)-(d) we consider $N=8$. 
Starting with the \sTSP{} and $n=N$, we: (a)
vary the number of observations $m$ while
keeping the total time $T = 2$; 
(b) fix the number of time steps
$m=10^4$ while varying the total time; and (c) fix the value of
$\delta t=5\times 10^{-4}$ while changing the number of time steps (and thus the
total time). For the \rTSP{}, we also (d) show how changing the number of time steps 
alters the diversity. Curves are averages over 100 realizations.
%
%
%
%
Panels (e) and (f) respectively show the rank diversity for the \sTSP{} when $n$ sites are allowed to move, and for
the \rTSP{} when $n$ sites are reallocated at each time step, both with  $N=7$
and $m=1000$. For the \sTSP{} the total time is $T=10$, which leads to $\delta
t=0.1$. We choose this value of $T$ so that (e) and (f) resemble
each other. Results in (f) are averages over 1000 realizations. 
Horizontal lines correspond to $d_{\max}$, the 
maximum value of the diversity given in \eref{eq:dmax}. In (b), (d), (e) and (f)
all curves have the same bound, whereas in (a) and (c) color is used to indicate the bounds of each curve.
}
\label{fig:varyparameters}
\end{figure}

In order
to quantify the way in which trajectory ranks change in time, we
shall use the rank diversity.
Moreover, to develop some intuition on this measure and understand how it depends on the
parameters chosen to calculate it, we present several numerical experiments for both models in 
\fref{fig:varyparameters}. 
Note that to calculate the rank diversity we have to choose an appropriate time
span $T$ and a number $m$ of time points at which observations are made.
 Alternatively, the time interval $\delta t$ between
observations can be chosen instead of $m$, where $T=m \delta t$.
%
%
In many real-world systems the time interval $\delta t$ is determined by data availability; to calculate the diversity
of words in English throughout the centuries, for example, one may use Google's n-gram dataset,
which implies a time interval $\delta t$ of one year (or an integer multiple of that). 
However, in the bTSP both the total time $T$ and 
the time interval $\delta t$ can be chosen at will.

Let us analyze the situations depicted in \fref{fig:varyparameters}.
First, we study the bTSP with a fixed total time evolution $T$ and varying
$m$,
thus effectively changing the value of $\delta t = T/m $ [\fref{fig:varyparameters}(a)]. 
As $m$ increases, the time between observations decreases, and the boats move
less, so the positions of the boats barely change in a single time interval
$\delta t$, and rank diversity diminishes. 
%
%
%
In fact, the diversity cannot be larger than the maximum
number of trajectories divided by the number of time slots,
$k_{\max}/m$.  If, on the other hand, $m$ is small enough, $k_{\max}/m$ is
larger than 1 and $d(k)$ tends
to saturate around 1. This can be summarized in the formula
\begin{equation}
d_{\max} = \min\left\{1, \frac{(N-1)!}{2m}\right\}.
\label{eq:dmax}
\end{equation}
In the second scenario, we fix the number of observations and vary the total time
$T$. As a limiting case, when $T$ is much larger than the time a boat needs to travel 
from one wall to another, the correlation between site configurations is
lost, so for most ranks diversity is close to its maximum; only 
close to the shortest and longest trajectories we observe smaller diversities [\fref{fig:varyparameters}(b)]. Since both $N$ and $m$ are
fixed, in this case $d_{\max}$ is the same for all $T$ values. 
We also analyze the case of a fixed value of $\delta t$ .
The value of $m$ now changes, as in \fref{fig:varyparameters}(a), but
simultaneously  
 $T$ varies, as in \fref{fig:varyparameters}(b), due to the relation 
$\delta t = T/m $. As $m$ increases, the effect on diversity seems 
similar to the one in which $T$ is constant [\fref{fig:varyparameters}(c)]. Finally, we perform a similar analysis for the rTSP [\fref{fig:varyparameters}(d)]. 
As $m$ increases, the time between observations is not shortened. Each new position 
of the selected site is uncorrelated with the previous one, regardless of the
value of $m$. We thus expect and observe a similar effect as the one of 
\fref{fig:varyparameters}(c). In this case, the bound for different values of $m$ does
not change, since $(N-1)!/2m >1$ for all cases shown. 
\par

%


Two useful generalizations of the
bTSP and the rTSP are now considered, since it will allow us
to relate the behavior of both models.
Assume that only some sites move in the bTSP. 
That is, instead of all 
site moving in the plane, $n$ move and
$N-n$ are static. The cases analyzed before thus correspond to $n=N$. 
In a similar way, consider an rTSP where instead of reallocating 
a single site, $n$ sites are moved. The case $n=N$ thus corresponds
to a total reallocation of the system, while up to now we have only discussed the
value $n=1$. 
Diversity for these generalizations
seems to behave in a similar way as for the case $n=1$, as seen in
\fref{fig:varyparameters}(e) and \fref{fig:varyparameters}(f).
In fact, for the bTSP one may even consider a
single moving site and 
the previous conclusions still hold.  
We have obtained similar results for other variations of the bTSP, such as
periodic boundary conditions, and different ways of choosing the initial conditions and
velocities.

\begin{figure} 
  \begin{center}
    \includegraphics{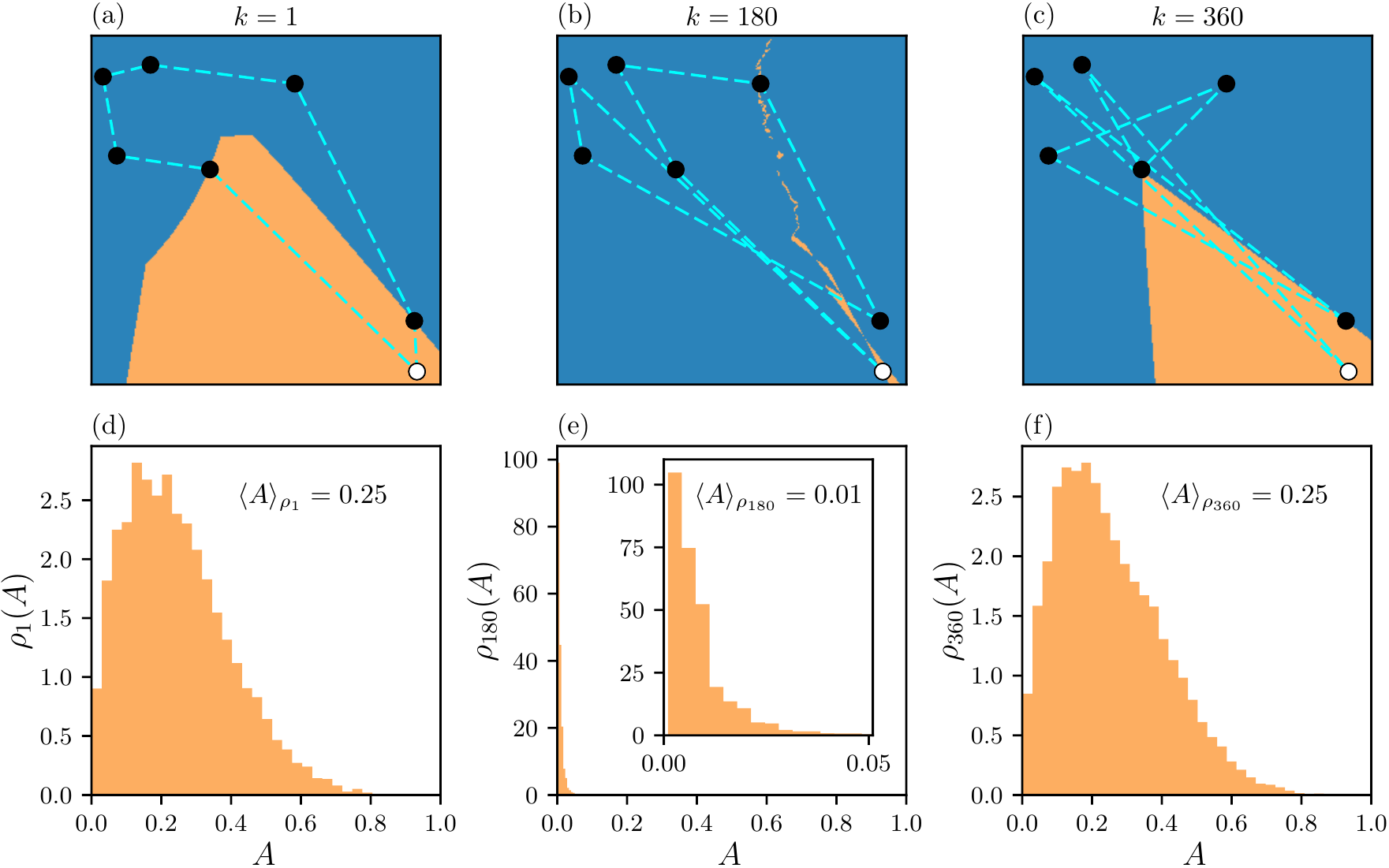}
  \end{center}
  \caption{ 
    {\bf Stability of ranks under reallocation of sites.}
    (a) Particular instance of the rTSP with $N=7$ and $n=1$. Black points 
represent fixed sites, whereas the white one represents the site
to be reallocated. The shortest trajectory is indicated as a dashed red line.  When the white
point is reallocated in the green area $A$, the shortest
trajectory remains unchanged, while if a node is reallocated in any part of the blue
area, the shortest trajectory changes. (b) Same realization of the rTSP, but for the trajectory with an intermediate rank
($k=180$). The green area is much smaller in this case. (c) Same realization of the rTSP, but for the longest
trajectory ($k=360$), where the green area is again big. In panels (d), (e) and (f) we show the
probability distributions of $A$ for $N=7$ and the three previous values of $k$, as well as its mean 
value, $\langle A \rangle_{\rho_k}$.
  }
  \label{fig:creation:annhilation}
\end{figure}

Let us analyze geometrically a particular, but random, instance of the rTSP 
with the goal of gaining some insight.
The configuration
space of such problem has dimension $2n$, as each of the $n$ sites has a two dimensional
space to move. For a given initial configuration, in
the case $n=1$ we can plot the locus of positions to which a given point may
travel, so the trajectory associated with a given rank remains unchanged.
In Figures~\ref{fig:creation:annhilation}(a-c) we take $N=7$ as an example, so the shortest,
middle and longest trajectories correspond to $k=1, k=180$, and $k=360$, respectively. The
example is representative in the sense that, for the extremal trajectories, the
aforementioned locus has a big area and the middle one has a very small area, but
particular values change broadly from realization to realization. We can also
  obtain a histogram of the probability density $\rho_k$ of the areas
  corresponding to random realizations
[\fref{fig:creation:annhilation}(d-f)]. From these histograms we see
that $\rho_1 \approx \rho_{k_{\max}}$, but $\rho_1$ is clearly different from
$\rho_{k_{\max}/2}$.

\begin{figure} 
  \begin{center}
    \includegraphics{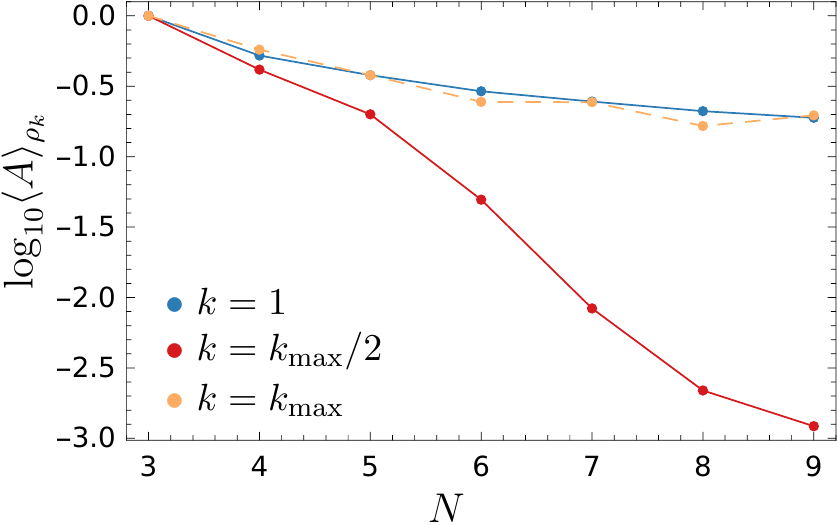}
  \end{center}
  \caption{{\bf Scaling in rank stability.} Scaling of the average value of the 
  area for which a change in position will not change the rank of 
  a trajectory, for ranks $k=1$, $k_{\max}/2$ and $k_{\max}$, and for 
  different system sizes.  
  }
  \label{fig:scaling}
\end{figure}

It is also useful to study the expected value 
of the stability area for different ranks, $\langle A \rangle_{\rho_k}$, 
as a function of the number of sites (\fref{fig:scaling}). 
There is a different scaling for the extremal and intermediate
rankings, so we expect that the difference in areas seen for the case $N=7$ is exponentially 
larger for bigger systems.  Although the standard deviation of the distribution $\rho_k$ is
relatively
large, the average value $\langle A \rangle_{\rho_k}$ 
can be determined with high accuracy (the statistical error for points
in \fref{fig:scaling} are comparable to the size of the points). 
We can also see an even-odd effect for the longest trajectory associated
with the change in topology of a star-like configuration for an even or 
odd number of sites. This hints at a
geometric understanding of time-dependent TSPs which, however, is outside the scope of
the present study.




\begin{figure} 
  \begin{center}
    \includegraphics{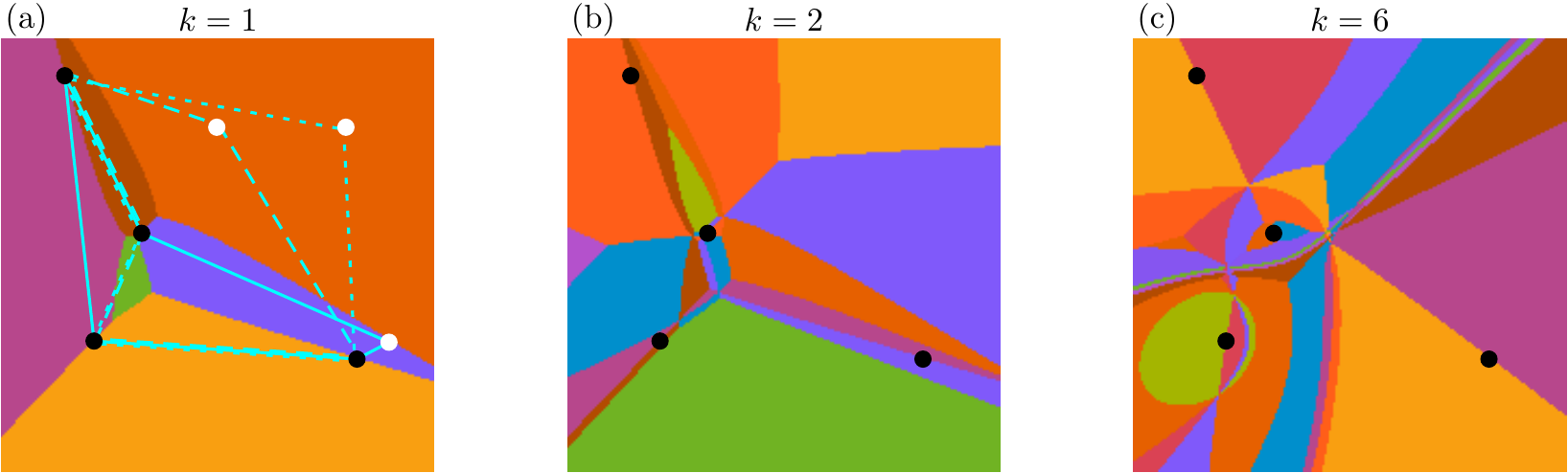}
  \end{center}
  \caption{ 
    {\bf Stability areas for different trajectories.}
(a) rTSP with $N=5$, $n=1$,
and three possible reallocations of the white point. 
The shortest trajectory, $k=1$, is
indicated as a solid, dashed and dotted line, corresponding to the three reallocations. 
Regions of varying color correspond to different trajectories for $k=1$ (see
\fref{fig:creation:annhilation}, top row). In panels (b) and (c) we show a plot with the same fixed sites as in (a), 
but for $k=2$ and $k=6$, respectively. 
  }
  \label{fig:psico}
\end{figure} 

\subsection*{Connecting the bTSP and rTSP} 

Consider the rTSP with $n=1$.   For a particular rank $k$, each position in the unit 
square yields a trajectory. We can thus `paint' the unit square
with colors corresponding to trajectories. We show examples with $N=5$ for the
shortest trajectory in \fref{fig:psico}(a), and for $k=2$ and
$6$ in \fref{fig:psico}(b) and \fref{fig:psico}(c), respectively. Diversity 
is the number of different colors in the picture divided by the number of
observations. Notice that there is a qualitative difference between the 
cases $k=1$ and $k=6$. This indicates already that $d(1) < d(6)$ if 
a large ensemble is taken (so that errors due to finite sampling 
are small enough). 

%
%

For the bTSP with $n=1$, the moving boat will cover the whole unit square
uniformly for most choices of the velocities. The condition for having a
uniform covering is that the vertical and horizontal speeds are incommensurable,
which always holds for this model. 
%
When this occurs, time averages yield the same value
as space
averages, i.e., the system is {\it ergodic}~\cite{yaglom1962stationary}.  
For such long times the whole unit square will be visited, i.e. the moving
site will visit all colored areas. Therefore, the bTSP has the  same
rank diversity as the rTSP when the sampling $m$ is the same (otherwise they are
related by a constant factor).

For a larger number of moving sites, $n>1$, a similar reasoning 
holds. However, the configuration space is now $2n$ dimensional, and thus
visualizing it  is more challenging. The condition for ergodicity 
still holds when all pairs of 
velocities are incommensurable. However, when the number of boats that move is increased,
the diversity changes as in \fref{fig:varyparameters}(e) (where three different values of
$n$ are shown). When $n=N-1$ or $N$, the diversity is 
formally maximum and we obtain no relevant information. In fact, 
when the time over which we calculate the diversity in the bTSP increases, 
$d(k)$ tends to 1 [\fref{fig:varyparameters}(b)]. 

The bTSP and rTSP are
comparable since both explore a fraction of the $2N$-dimensional configuration
space. The bTSP explores a line of finite length 
in such a space, or a $2N$
dimensional hypercube embedded in the configuration
space if the whole ensemble of velocities
is considered. The rTSP, on the other hand, explores a $2n$ dimensional 
hyperplane embedded in the same configuration space. Overall, 
both models behave in a similar fashion.


Each realization of the TSP can be seen as a point in a $2N$-dimensional configuration space, 
where every pair of axis defines the coordinates of each particle. The optimization 
problem is then different for each point in the configuration space. 
%
%
We have analyzed the stability of the solutions of the TSP under changes of the location of the point
defining the configuration.
We have further shown that the stability properties are similar for the two time-dependent generalizations of the TSP
considered here. We have also stated under what conditions the behavior of both models is identical. We thus expect that
these results are applicable to other perturbations of the TSP that involve small variations 
in configuration space.

\section*{Discussion} 

We have studied the statistical properties of rank distributions and rank dynamics of a
novel variation of the traveling salesman problem where nodes shift their position in
time. This allows us to explore the stability of trajectory ranking, which is related to
the predictability of  perturbation effects. \fref{fig:scaling} shows the average
probability of a trajectory maintaining its rank of 1000 instances of the bTSP (with one
site moving) as the number $N$ of sites increases. We see that this probability,
which reflects the stability of ranks to perturbations, decreases with $N$. However, the
decrease is much lower for the shortest and longest paths than for
the middle trajectories. This reflects the fact that the rank diversity is also lowest for 
the extreme paths. Moreover, the stability decreases much faster for the
middle trajectories. Thus, we find that the shortest and longest trajectories are
more predictable and robust. This claim would benefit from analytic solutions for the numerical results presented here, but these are beyond the scope of the present work.

We also note that rank diversity curves are all symmetric, as shown in
\fref{fig:varyparameters}. However, in previous studies we have
shown that the rank diversity $d(k)$ has almost the same form for languages, sports and
games. This is not symmetric and can be adjusted by a sigmoid in lognormal scale. Still,
from other datasets we have also noticed symmetric rank diversity curves. The difference between symmetric and
non-symmetric rank diversity curves seems to be related with the degree of `openness' of a
system. In time-dependent TSPs the system is `closed', as all trajectories are
considered at all times. However, in sports, languages and other real-world phenomena, elements enter and exit the
system. More precisely, elements enter and exit the subset of the system available in datasets. If
one plots rank diversity curves of `open' systems in linear scale, they are very similar
to the symmetric curves of closed systems~\cite{cocho:idiomas,cochodos}.


Changes in optimization problems pose challenges when change
is faster than optimization, as solutions might be
obsolete~\cite{Gershenson2013Facing-Complexi}. These non-stationary problems
are a common feature of complex systems~\cite{Gershenson:2011e}. Our analysis
suggests that the way in which state spaces of non-stationary problems change
is not uniform. This implies that once an optimal solution is found, we can expect it to be more stable than non-optimal solutions. A more
detailed exploration of the changes of state spaces in time will provide
further insight, and we trust that the methods presented here will
contribute to this effort.

%
%
%

\section*{Data Availability} 
The datasets generated during and/or analyzed during the current study are available from the corresponding author on reasonable request.
\bibliography{references}
\section*{Acknowledgements} 
We acknowledge support from UNAM-PAPIIT Grant No.
IN111015.
\section*{Author contributions statement} 
J.F. and C.P. conceived the ideas behind this work. G.C. obtained the rank distribution for
the static TSP. S.S. and C.P. carried out the numerical simulations and statistical
tests. G.I. provided the framework to contextualize the results. J.F., C.P., G.I., and C.G. wrote the manuscript. All authors discussed and reviewed the manuscript. 
\section*{Additional information} 

\textbf{Competing financial interests}\\
The authors declare that they have no competing interests.
\end{document}